\documentclass[review]{elsarticle}

\usepackage{amssymb}
\usepackage{graphicx}
\usepackage{times}
\usepackage{amsmath}
\usepackage{amsfonts}
\usepackage{textcomp}
\usepackage{multirow}
\usepackage{array}
\usepackage{url}
\usepackage[usenames,dvipsnames]{color}
\usepackage{float}
\usepackage{subfigure}
\usepackage{color}
\usepackage{pgfplots}
\usepackage{fancyhdr}
\usepackage{siunitx} 
\usepackage{booktabs}
\usepackage{graphicx}

\journal{Journal of Extreme Mechanics Letter}

\begin{document}

\begin{frontmatter}

\title{Tailorable Elasticity of Cantilever Using Spatio-Angular Functionally Graded Biomimetic Scales }

\author[mymainaddress]{Hessein Ali}
\author[mymainaddress]{Hossein Ebrahimi}
\author[mymainaddress]{Ranajay Ghosh\corref{mycorrespondingauthor}}
\cortext[mycorrespondingauthor]{Corresponding author}
\ead{ranajay.ghosh@ucf.edu}

\address[mymainaddress]{Department of Mechanical and Aerospace Engineering, University of Central Florida, Orlando FL, 32816}

\begin{abstract}
Cantilevered beams are of immense importance as structural and sensorial members for a number of applications. Endowing tailorable elasticity can have wide ranging engineering ramification. Such tailorability could be possible using some type of spatial gradation in the beam’s material or cross section. However, these often require extensive additive and subtractive material processing or specialized casts. In this letter, we demonstrate an alternative bio inspired mechanical pathway, which is based on exploiting the nonlinearity that would arise from a functionally graded distribution of biomimetic scales on the surface using an analytical approach. This functional gradation is geometrically sourced and could arise from either spatial or angular gradation of scales.  We analyze such a functionally graded cantilever beam under uniform loading. In comparison with uniformly distributed scales, we find significant differences in bending stiffness for both spatial and angular gradations. Spatial and angular functional gradation share some universality but also sharp contrasts in their effect on the underlying beam. A combination of both types of gradation in the structure can be used to alternatively increase or decrease stiffness and therefore a pathway to tailor the elasticity of a cantilever beam relatively easily. These results give rise to an architected framework for designing and optimizing the topography of leveraged solids.
\end{abstract}

\begin{keyword}
Bio inspired \sep cantilever \sep functionally graded material (FGM) \sep tunable elasticity
\end{keyword}

\end{frontmatter}


Cantilevered beams arise in a number of diverse engineering applications spanning an enormous variety of length scales~\cite{Can1,Can2,Can3,Can4,Can5,Can6,Can7,Can8,Can9,Can10}. In several of these applications, tailoring elasticity can be of tremendous significance since it can be used to design the response according to stimulus or guard against unwanted instabilities. To this end, functional gradation (FG) is a useful strategy. FG materials are high performance composite materials consisting of two or more constituent phases with variegation in composition. This gradation can lead to a desired enhancement in the thermal/mechanical properties, compared to their conventional counterparts. This makes them ideal for various engineering applications including biomedical~\cite{FGbio1,FGbio2}, cellular structures~\cite{FGstr1,FGstr2,FGstr3,FGstr4,FGstr5,FGstr6}, soft robotics~\cite{FGrob1,FGrob2} and several others~\cite{FGoth1,FGoth2}.

However, typical FG materials could be difficult to fabricate requiring extensive materials processing such as directional solidification [23], specialized machining paths or even additive manufacturing~\cite{FGman1,FGman2}. An alternative exists in pursuing surface based strategy such as biomimetic scales. Such dermal scales are a pervasive feature within Kingdom Animalia. Their advantages extend well into a variety of important functions, which enhance survivability, such as protection, camouflaging, and locomotion~\cite{Scalesfun1,Scalesfun2,Scalesfun3,Scalesfun4,Scalesfun5}.  In nature, certain fishes possess remarkably periodic scale distribution, for instance, \textit{Elasmobranchs}~\cite{Elas1,Elas2} and \textit{Teleosts}~\cite{Tele1,Tele2,Tele3,Tele4}.  However, more often, organisms display a large variation in scale distribution within their own bodies~\cite{Variation1,Variation2,Variation3,Variation4}. This is primarily due to both physiological factors~\cite{Physfac1,Physfac2,Physfac3} and in response to the functional requirements~\cite{Funreq1,Funreq2,Funreq3} resulting in varying density of scales and even their arrangements. Members of the order \textit{Squamata} and \textit{Crocodilia} display axial functional scale gradation, to improve locomotion~\cite{Funreq1}. Their scale density increases near the head and tail in order to enable a better spatial mobility~\cite{Funreq1}. This gradation also varies among species, for instance in snakes, boas, pythons, and many vipers have small, irregularly-arranged head scales in contrast to the large systematic head scales of most advanced snakes~\cite{Funreq1}. Interestingly, scaly features with functional gradation are also common in hair and furs of mammals~\cite{Furfun}. 

At the laboratory scale, scaly structures can be fabricated by embedding plate-like structures into a soft substrate~\cite{Sfabr1,Ghosh1,Sfabr2}. Extensive work has been conducted on the kinematics and mechanics of scaly structures from the point of view of localized loading such as armor applications~\cite{Armapp} as well as fostering nonlinearities in global deformations in bending and twisting stemming from scales engagement~\cite{Ghosh1,Frank1,Frank2,Frank3,fishmech1,Ghosh2,Hossein1,Hessein1}. In these studies, the uniformly distributed scales are considered. This considerably simplifies the design and analysis but unfortunately also preclude the many advantages that come from functional gradation. 

\begin{figure}[ht!]
\centering
      \subfigure[]{
        \includegraphics[width=0.48\columnwidth]{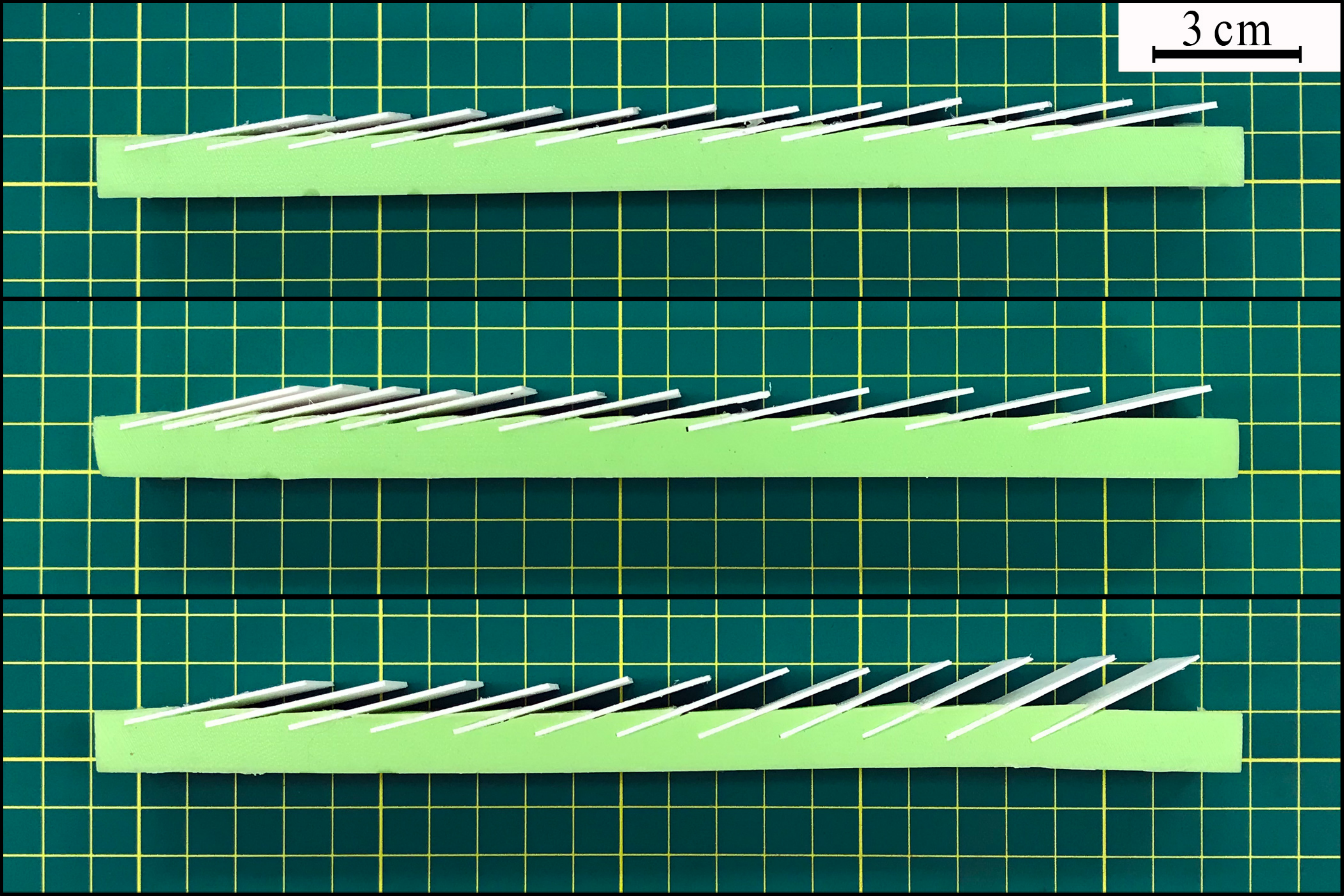}
        \label{fig:FG1A}}
				\quad
				\hspace{-3ex}
      \subfigure[]{
        \includegraphics[width=0.48\columnwidth]{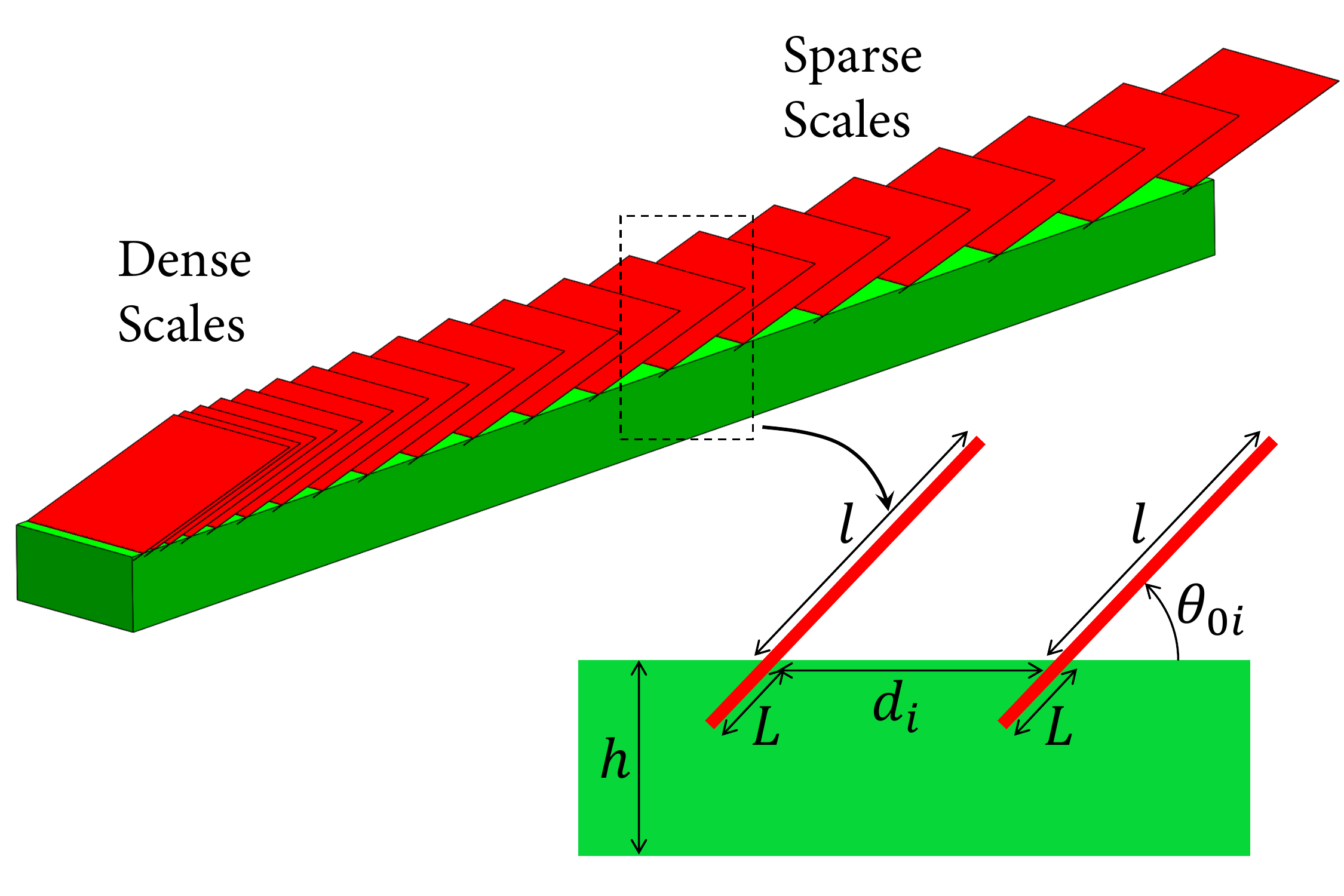}
        \label{fig:FG1B}}
    \subfigure[]{
      \includegraphics[width=0.48\columnwidth]{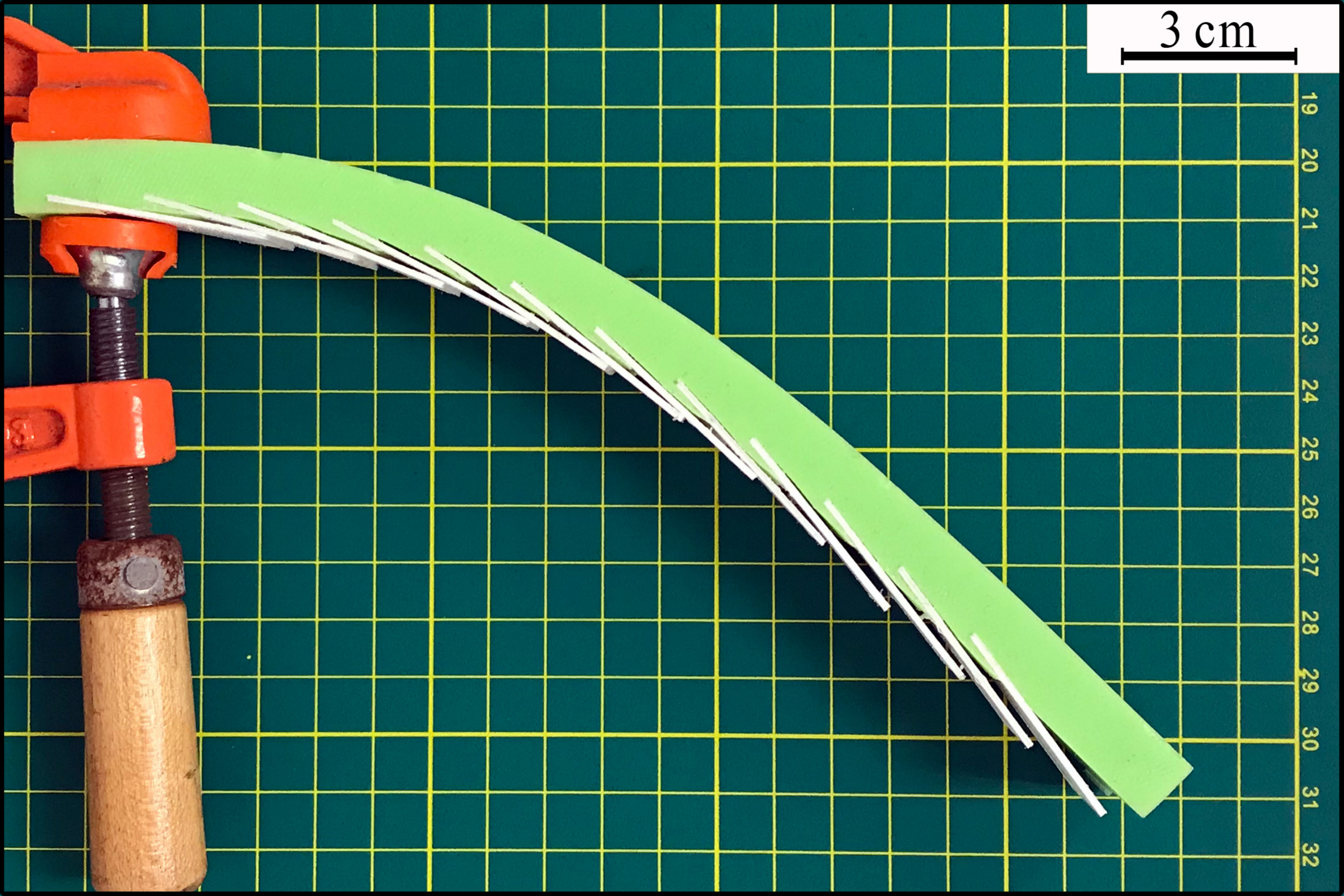}
      \label{fig:FG1C}}
			\quad
			\hspace{-3ex}
			\subfigure[]{
      \includegraphics[width=0.48\columnwidth]{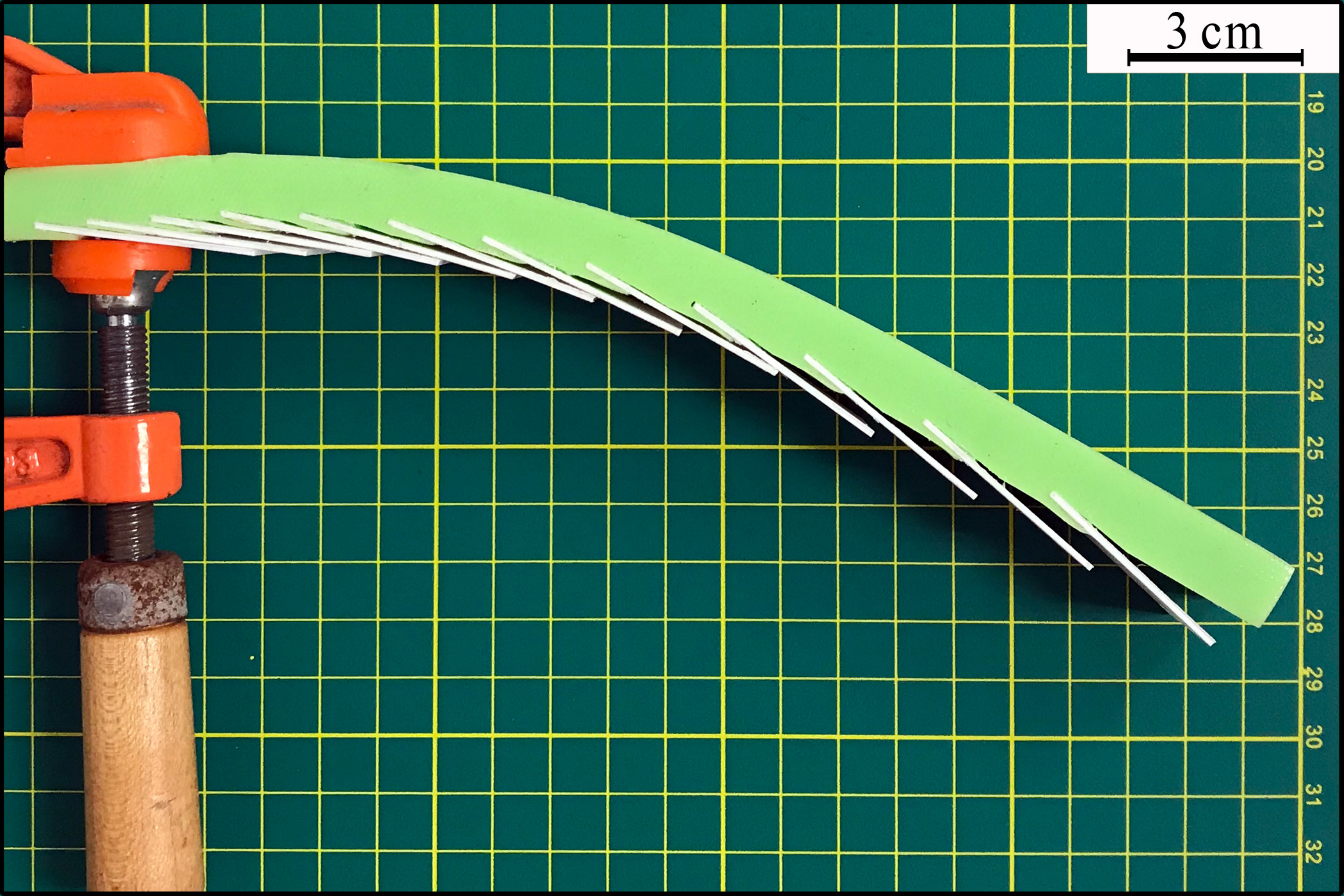}
      \label{fig:FG1D}}
	\caption{(a) Three scaly beams (from top to bottom) in which scales are uniformly distributed, linearly placed along the beam, oriented with linear gradation in the inclination angle. (b) A 3D SolidWorks model of a substrate with embedded scales arranged linearly along the length of the beam and a schematic diagram of two adjacent scales. (c) The deflection of a uniformly scaled beam due to beams self-weight. (d) The deflection of a beam with linearly arranged scales on its top surface with constant initial inclination angle.The dimensions of the beam and scales are $220$ mm (length) $\times$ $25$ mm (width) $\times$ $10$ mm (height) and $35$ mm $\times$ $25$ mm $\times$ $1$ mm, respectively.}
\end{figure}

In this letter, we study the effect of functional gradation in modulating the behavior of biomimetic scale covered cantilevered beam. Scales on the surface can demonstrate functional gradation in two distinct geometrical variables - spatial (scale distribution) and angular (scale inclination), Fig.~\ref{fig:FG1A}. We demonstrate the effect of these gains using qualitative experiments of beam deflection, Figs.~\ref{fig:FG1C} and ~\ref{fig:FG1D}. These figures illustrate a noticeable difference in deflection between two similar FG scaly samples, one with uniform distribution and the other linear FG, under their own self weight. These samples were fabricated through adhering $3D$ printed scales (Poly Lactic Acid (PLA)) with an elastomer known as Vinylpolysiloxane (VPS). The linearly arranged scaly sample demonstrates a decrease of $30 \%$ in deflection, compared to uniform scaly sample (a direct observation from the scaled mat). 

Motivated by these purely qualitative experiments, we investigate the behavior in detail using a combination of analytical models aided by computational investigations~\cite{Hessein1}. The FG scaly beam is composed of an underlying substrate and partially embedded scales on its surface, Fig.~\ref{fig:FG1B}. The substrate is of length $L_B$ and height $h$, while each individual scale of thickness D has an exposed part of length $l$  and embedded portion $L$, which makes the total length of scale be $l_s=l+L$ . We refer to the overlapping ratio of scales as $\eta=l/d$, where $d$ is the spacing between two adjacent scales~\cite{Ghosh1}. Note that as we introduce spatial gradation,  $\eta$ will vary with position. In such FG beams, all scales would start from an initial inclination angle $\theta_0$ measured with respect to the beam centerline and increase after engagement in a highly nonlinear fashion~\cite{Ghosh1,Frank1,Hessein1}. We measured the Young’s modulus of the of VPS and PLA via tensile test using an MTS Insight\textregistered, and found them to be $1.5$ MPa and $2.86$ GPa, respectively.

This high contrast in the modulus along with the assumption $D \ll l_s$ and $L\ll h$ (Shallow embedding) allows us to model the scales as rigid plates embedded in a semi-infinite elastic media. Accordingly, the scales rotation is modeled as a linear torsional spring~\cite{Ghosh1,Frank1}. We further impose small strains in the beam, and hence Euler-Bernoulli assumptions remain applicable.

\begin{figure}[htbp]
\centering
\subfigure[]{%
\includegraphics[scale = .65]{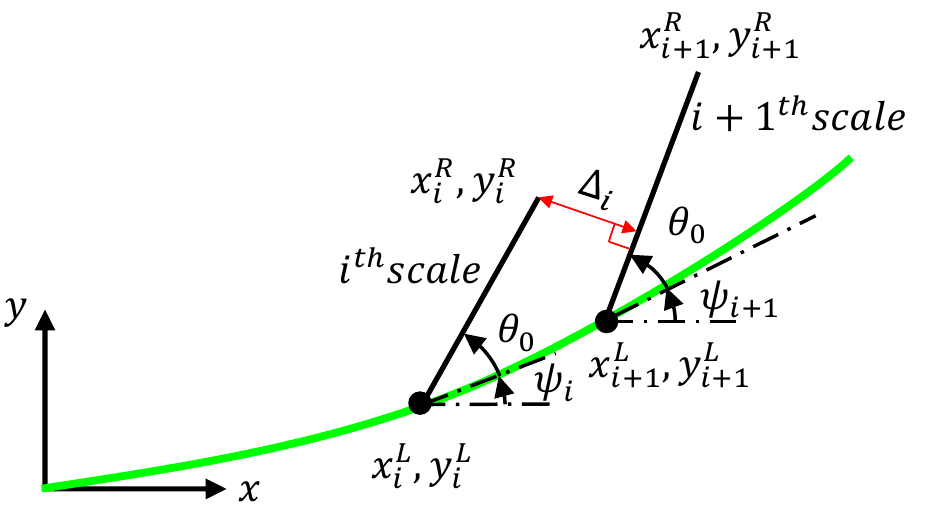}
\label{FG2A}}
\quad
\subfigure[]{%
\includegraphics[scale=0.65]{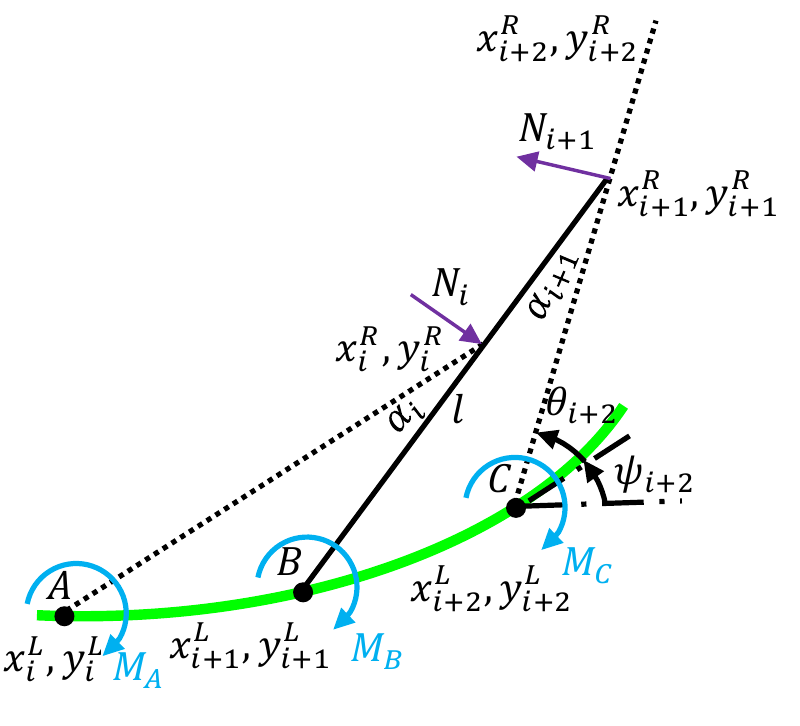}
\label{FG2B}}
\caption{(a) The geometry of scales before engagement at a configuration of a deformed underlying substrate. (b) The FBD of a scale when it is in contact with two neighboring scales shown as the dotted lines for clarity (adapted from~\cite{Hessein1}).}
\label{fig1} 
\end{figure}

We start our model by first acknowledging the lack of global periodicity in FG scaly systems due to the non-periodic engagement of scales and non-uniformity of bending caused by the boundary and loading conditions~\cite{Hessein1}. This prevents us from using the same form of kinematic relationship developed earlier~\cite{Ghosh1,Frank1}. Therefore, we utilize the approach of discrete scales~\cite{Hessein1} to address the kinematic of scales in our cantilever FG scaly beam. We impose a deformation on the beam in the form $y(x)=\gamma f(x)$ , where $f(x)$ is a shape function and  $\gamma$ is a dimensionless constant determined by the beam material, geometry, and load. The engagement of scales can then be tracked by means of the distance parameter $\Delta_i$ of the right extremity of each scale and its nearest subsequent neighbor, Fig.~\ref{FG2A}. This distance parameter can be written as~\cite{Hessein1}:
\begin{equation}
\Delta_i= {1 \over l}((y_{i + 1}^L - y_{i + 1}^R)(x_{i + 1}^L - x_i^R) - (x_{i + 1}^L - x_{i + 1}^R)(y_{i + 1}^L - y_i^R)),i = 1,..,{N_s} - 1
\label{eq2}
\end{equation}

where $N_s$ is the total number of scales. The engagement occurs when $\Delta_i \le 0$ where $i=1,...,N_e$ where $N_e$  would be the number of scales in contact. This results in a set of $3N_e-2$ nonlinear equations after the condition of scales interaction is met.  These equations consist of constraints on the fixed length of scales, the geometry of engagement, and balance of moments about the base of all engaged scales, Fig.~\ref{FG2B} (See Supplemental Material for derivation~\cite{sub}). The nonlinear equations are then solved numerically to ensure equilibrium at each step of deformation. The outcome of solving these equations is the orientation of each scale as the underlying substrate progressively deforms into an arc.

The resultant bending mode of FG system can be envisioned as a combination of substrate deformation and scales rotation. The strain energy due to the deformation of the underlying substrate can be written as  $\Omega_B= \int_0^{L_B} {1 \over 2} EI\kappa^2 dx$, where $EI$ is the bending rigidity of the beam and $\kappa$ is the instantaneous curvature. The scales energy is modeled as $\Omega_{scales}=\sum_{i=1}^{N_e}{1\over2} K_B (\theta_i-\theta_0 )^2$. Here $\theta_i$ is the scales rotational displacement evaluated using the kinematic approach of discrete scales (See Supplemental Material for how it is calculated~\cite{sub}), $N_e$ is the number of scales in contact, and $K_B$ is the spring constant of the linear torsional spring (rigid scales rotation) which models the resistance of the substrate to rotation of the embedded scale. The stiffness $K_B$ is analytically approximated as $K_B=C_B ED^2 (L/D)^n$ where $E$ is the modulus of elasticity of the substrate and $C_B,n$ are constants with values $0.86,1.75$, respectively~\cite{Ghosh1,Hessein1}. The total potential energy is $\Pi=\Omega_{beam}+ \Omega_{scales} H(\Delta_i)-W$ where $W$ is the external work and  $H(\Delta_i)$ is the Heaviside step function. Note that the FG scaly beam will deflect in a shape similar to that of a plain beam, leading us to $y(x)=\gamma f(x)$ where $f(x)$ is the shape of the corresponding plain beam. This allows us to minimize the potential energy through variation with respect to $\gamma$.  In the differential form, the variation in energy can now be written as ${d\Omega_{beam}\over d\gamma}+{d\Omega_{scales}\over d\gamma} H(\Delta_i )={dW \over d\gamma}$, which allows us to compute the beam deflection under given load.  We first study the effect of beam deflection under its own self-weight. This can be modeled as a uniformly distributed load $w_0$ across its span. For this case, the deformation of the plain beam is $y(x)={{w_0 h^3}\over{24EI}} (4Lx^3-x^4-6L_B^2 x^2 )$~\cite{shigley1} , while the work done is  $W = \;\mathop \smallint \limits_0^{L_B} {w_o}y\left( x \right)dx$.  Once biomimetic scales start interacting during deformation, they add stiffness to the structure, which would require an additional amount of load to obtain an equivalent deflection similar to a plain beam. This equivalent load is derived utilizing the variational-energetic equation and can be written as:
\begin{equation}
w = {w_0} + {5h^3 \over {{6L^5}}}\mathop \sum \limits_{i = 1}^{{N_e}} {k_B}\left( {\theta_i  - {\theta _0}} \right){{d\theta } \over {d\gamma }}.
\label{eqFG2}
\end{equation}
Where $d\theta/d\gamma$  is numerically using finite differences. Note that the tip deflection of the beam will deviate from linearity due to the highly nonlinear regime brought about scales sliding~\cite{Ghosh1}. Equation~\ref{eqFG2} now serves as a proxy to track the tip deflection of the FG cantilever beam when scales sliding commences. For this communication, we fix the beam geometry parameters as  $L_B=1000$ mm and $h=50$ mm, while the scales parameters as $L=7$ mm,  $l_s=250$, $\theta_0=5^\circ$  , and $D=0.1$ mm. The modulus of elasticity of the beam is taken to be $E=1.5$ MPa.

\begin{figure}[ht!]
\centering
\subfigure[]{%
\includegraphics[scale = .6]{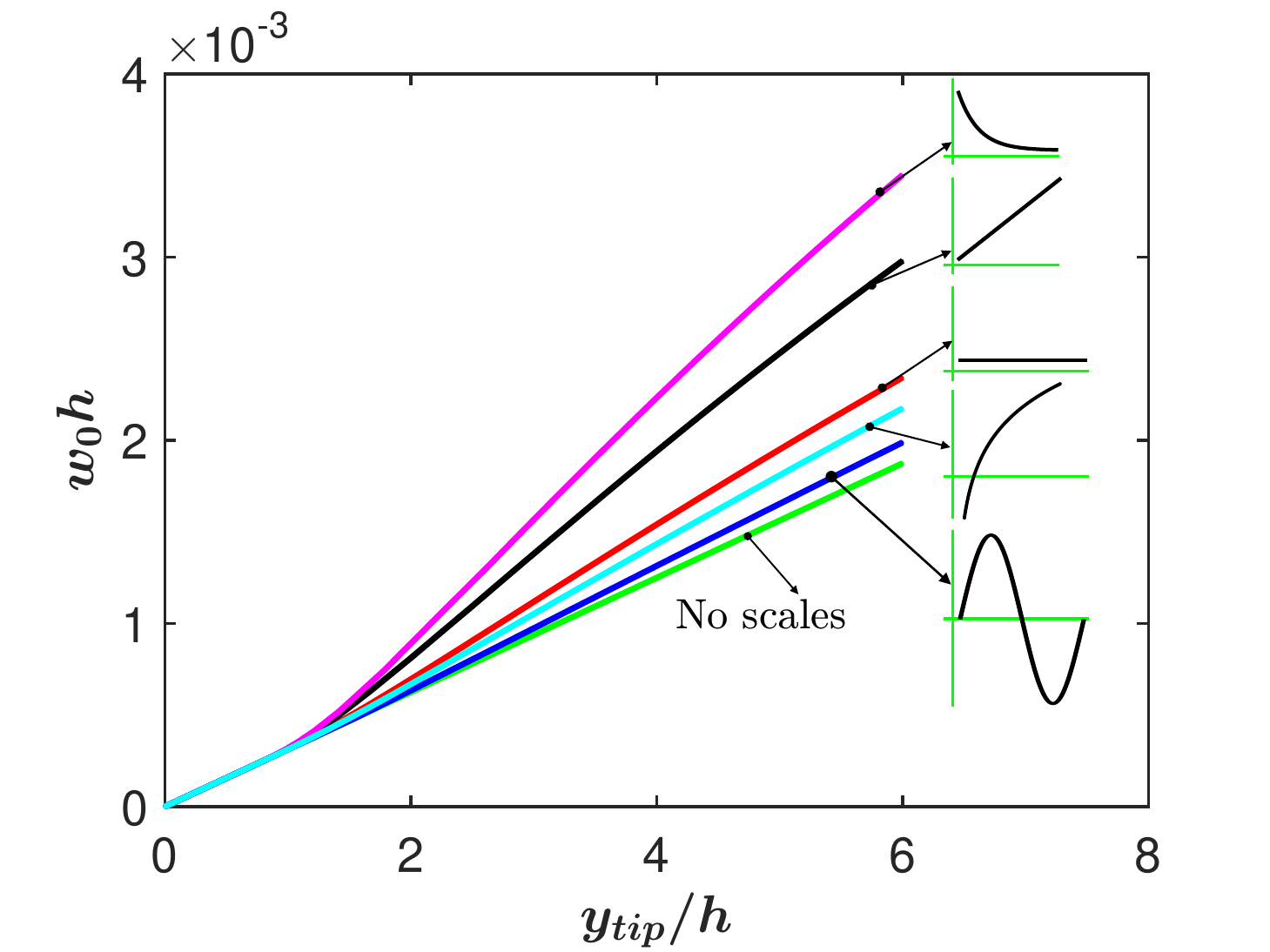}
\label{FG3A}}
\quad
\hspace{-6ex}
\subfigure[]{%
\includegraphics[scale=0.6]{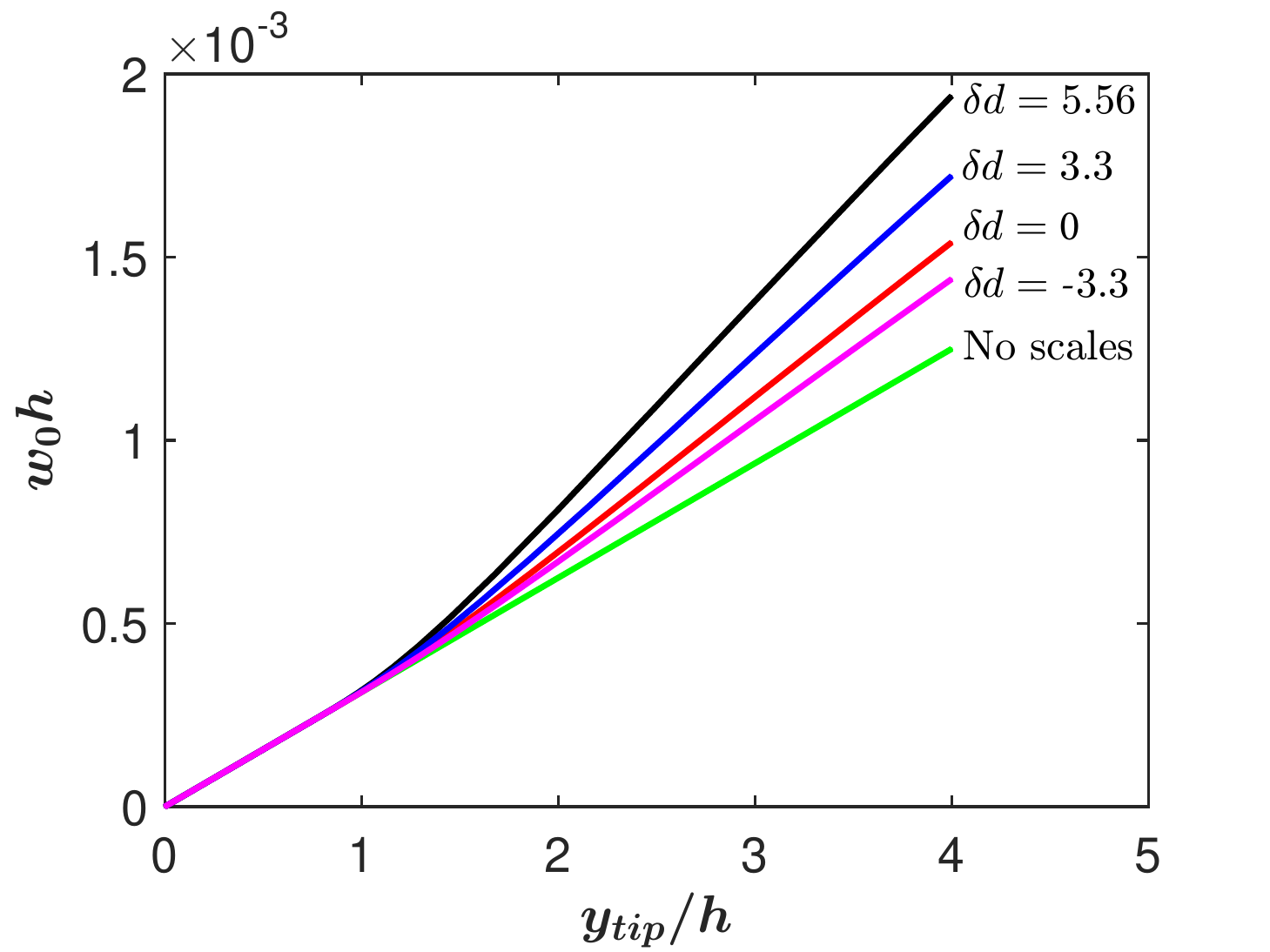}
\label{FG3B}}
\caption{(a) The load-displacement of FG cantilever scaly beam for a beam with $20$ scales that are functionally distributed where the density of scales is graphically represented. (b) The load-displacement for a beam with various linear gradation of scales.}
\label{fig1} 
\end{figure}

We first investigate the possibilities of functions that can be utilized for scales distribution along the length of our cantilever scaly beam.  These functions are then examined for the case of $20$ scales and fixed location of the first and last scale, ${x_o\over L_B} =0.01$ and ${x_f\over L_B} =0.98$. We start with an exponential distribution follows the formula $x=a\exp^{b s_n }$, where $s_n$ denotes the scale number while $a$ and $b$  are constants with values of $7.85$ and $0.24$, respectively.  We also implement a linear function in the form $x=x_0+d_0+\alpha(s_n-2)$ with $d_0=5$ mm and nondimensional $\alpha=5.11$. Another function of sparse distribution around the built in edge is  $x=x_0+a\log(s_n)$. Here the constant $a$ was found to be $323.8$. The last function tested is the case of having a beam of dense scales around its middle. The scales distribution is $x=a\sin{({{\pi s_n}\over {N_s}})}+b$ . where $a$ and $b$ are constants and found to be $569$ and  -$79$ , respectively. We plot the normalized tip deflection versus the normalized applied load for all tested functions, Fig.~\ref{FG3A} in which the legend of each curve is represented through the visualization of the function utilized (scales density). The results predict that exponential distribution of scales provides the potential of gaining the higher stiffness in the structure. However, we do not see a drastic difference in the results between the linear and exponential case unlike other possibilities. Therefore, linear gradation will be addressed in detail in this letter for brevity.

We begin with the case of spatial gradation. A linear-spatial gradation is imposed  in the form of  $d=d_0+\delta_d (s_n-2)$, $s_n>1$, where $d_0$ is the spacing between the first (left side of the beam) and the second scale, $\delta_d$ is the gradient and could be positive or negative, and $s_n$ is the scale number. A positive gradient would lead to more scales at the built-in side of the beam whereas the opposite is true for the negative. Furthermore, we assume a fixed number of total scales for the beam, $N_s=20$ along with the location of the first and last scale to uniquely determine the scale position along the beam for a given gradient in scale spacing. This therefore leaves $d_0$ as a function of $\delta_d$ allowing for a bigger parameter space compared to say exponential or sinusoidal. The plot of load-displacement characteristic is illustrated in Fig.~\ref{FG3B}. This figure shows that higher spatial gradation leads to both earlier engagement of scales as well as lesser deflection at same uniform loads. Particularly, for uniform scale distribution, scales engagement occurs when $w_0 h=2.6e-4$ , while a lower deformation is required for scales engagement, $w_0 h=2.25e-4$ when scales are spatially graded, $\delta_d=5.56$. Additionally, the imposition of linear gradation clearly expedites the instant of initial engagement of scales as compared to uniform scales. This quick engagement will lead to an earlier lock in the structure~\cite{Hessein1} as the beam progressively deforms and scales keep sliding until the point where they eventually stop. This local locking emerged due to the non-periodic engagement of scales, which forces the neighboring locked scales to undergo rigid body motion, while the rest of scales keep sliding~\cite{Hessein1}. Accordingly, the results of linear gradation open a way for the design of metamaterial structures that requires higher stiffness and quick local locking. 

\begin{figure}[htbp]
\centering
\subfigure[]{%
\includegraphics[scale = .41]{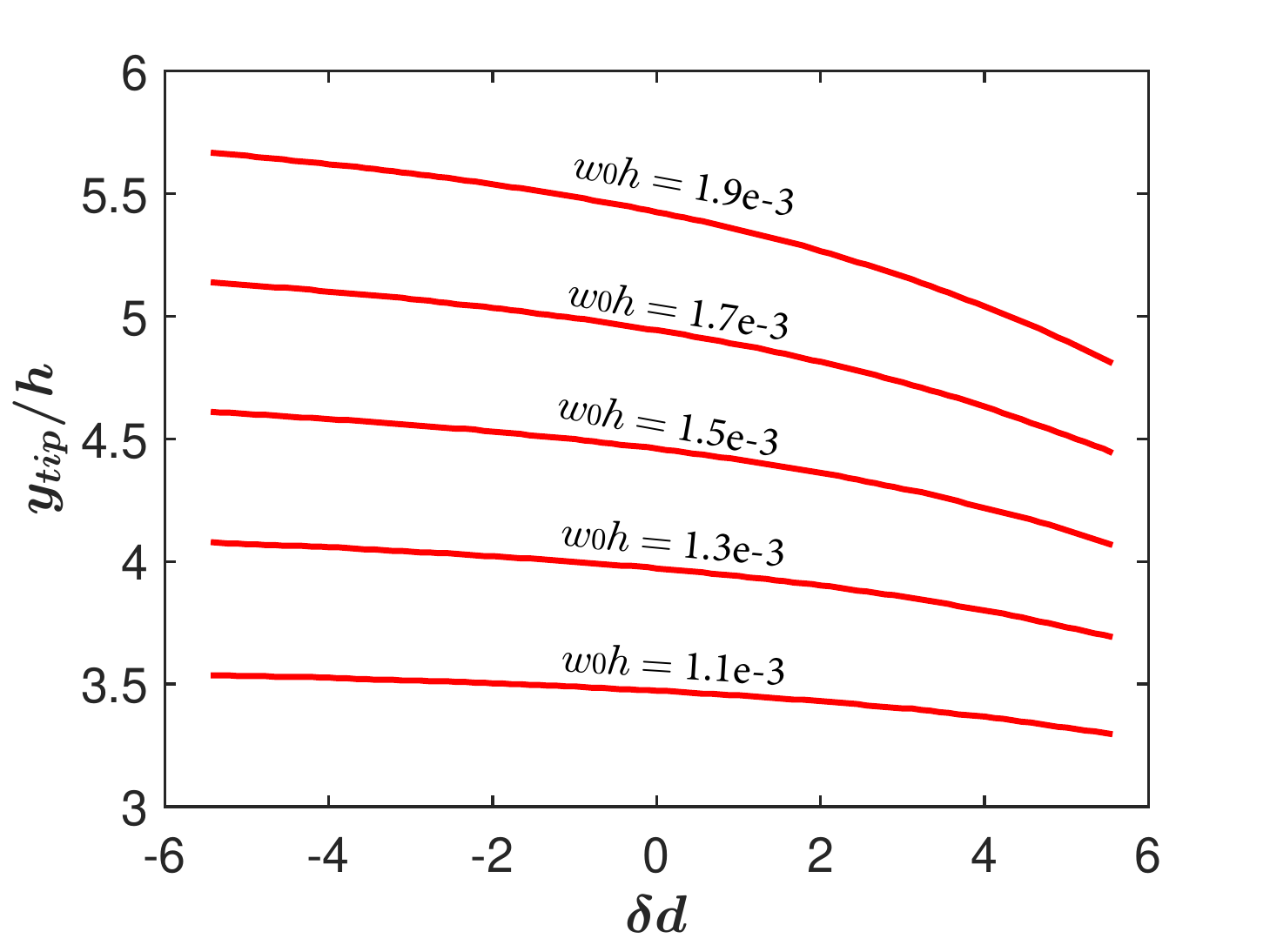}
\label{FG4A}}
\quad
\hspace{-6ex}
\subfigure[]{%
\includegraphics[scale=0.41]{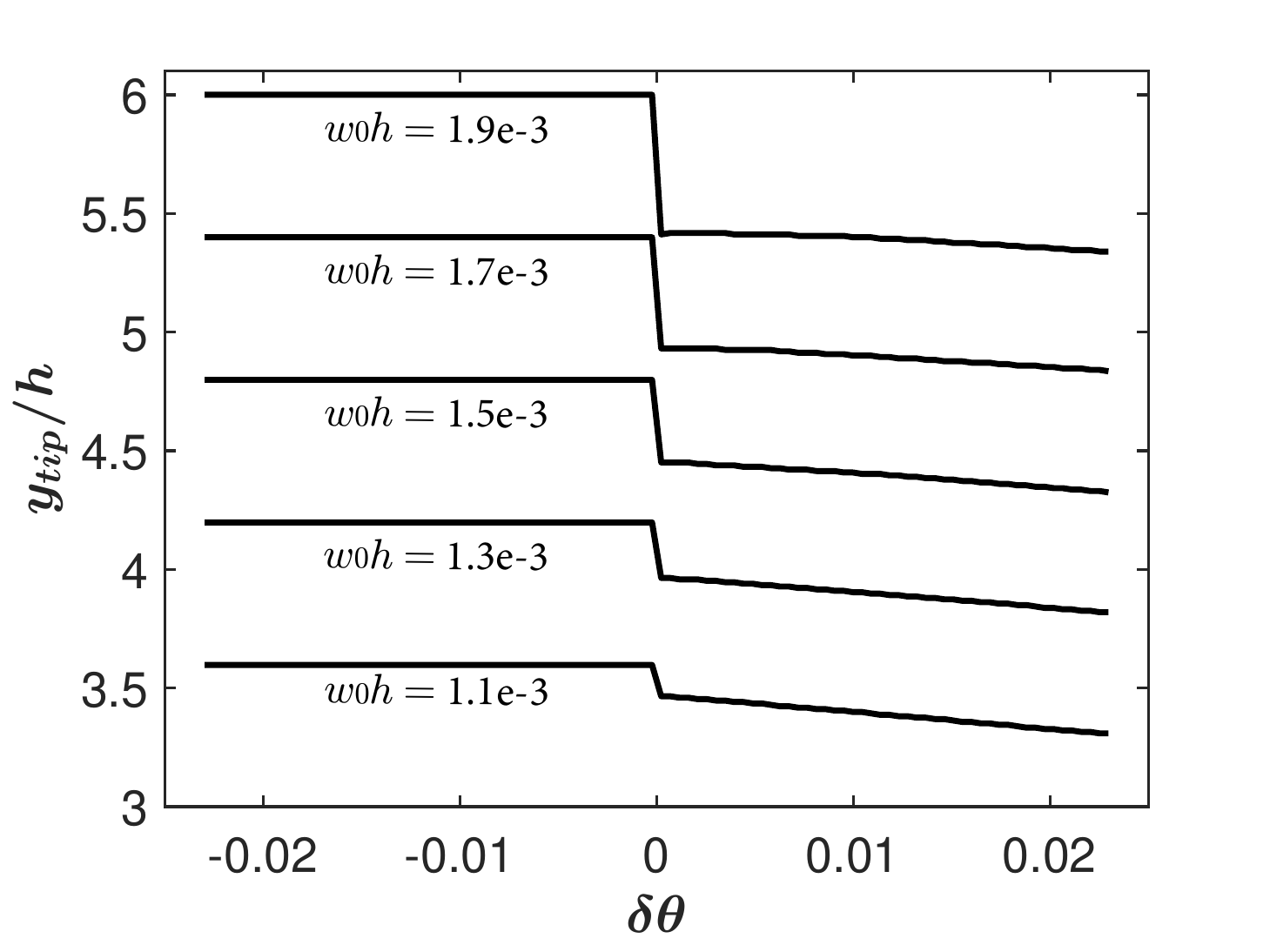}
\label{FG4B}}
\caption{(a) Non-dimensionalized tip deflection of FG cantilever scaly beam subject to distributed loading, $w_0$ for various gradient of spacing between scales. (b) The non-dimensionalized tip deflection of  uniformly loaded FG scaly beam with different initial inclination angle of each scale.}
\end{figure}

Interestingly, negative gradient, Fig.~\ref{FG3B}, delays the engagement of scales and increases the compliance when comparing with uniformly scaled beam. This motivates us to explore the efficacy of all possible gradients $[-\delta_{dmax},\delta_{dmax}]$ of scales for various applied loads. The normalized tip deflection is plotted with spatial gradient of scales for increasing load intensity in Fig.~\ref{FG4A}. Note that $\delta_d=0$ corresponds to uniform scales. This figure shows that higher spatial gradation leads to a decrease in tip deflection (increased stiffness) for positive gradients only. The trend is reversed for negative gradients. Therefore, placing more scales near the tip will lead to an increase in compliance when compared to uniformly distributed scales of same number. This positive-negative asymmetry is further sharpened at higher loads. This study shows significant tailorability of response through simple linear spatial functional gradation.

We now consider angular gradation effects on the beam. We again impose a linear variation of the initial inclination angle of scales. This is expressed through $\theta_0=\theta_{0i}+\delta_\theta(s_n-1)$ where $\theta_{0i}$ is the initial inclination angle of the first scale (again left side of the beam), $\delta_\theta$ is the gradient and could be positive or negative, and $s_n$ is the scale number. Here we address all possible choices of varying scales orientation in the range of $5^\circ \le \theta_0 \le 30^\circ$. Using the same dimensions of beam and scales as above, we plot the non-dimensional tip deflection versus $\delta_\theta $ for increasing load intensity in Fig.~\ref{FG4B}. Here, the positive-negative asymmetry is further accentuated since negative gradients will not lead to any engagement. However, unlike spatial gradient, here increasing load leads to more convergent stiffness across gradation. Therefore, angular and spatial functional gradation share some universality but also sharp contrast in their behavior. 

\begin{figure}[ht!]
\centering
\includegraphics[scale = .6]{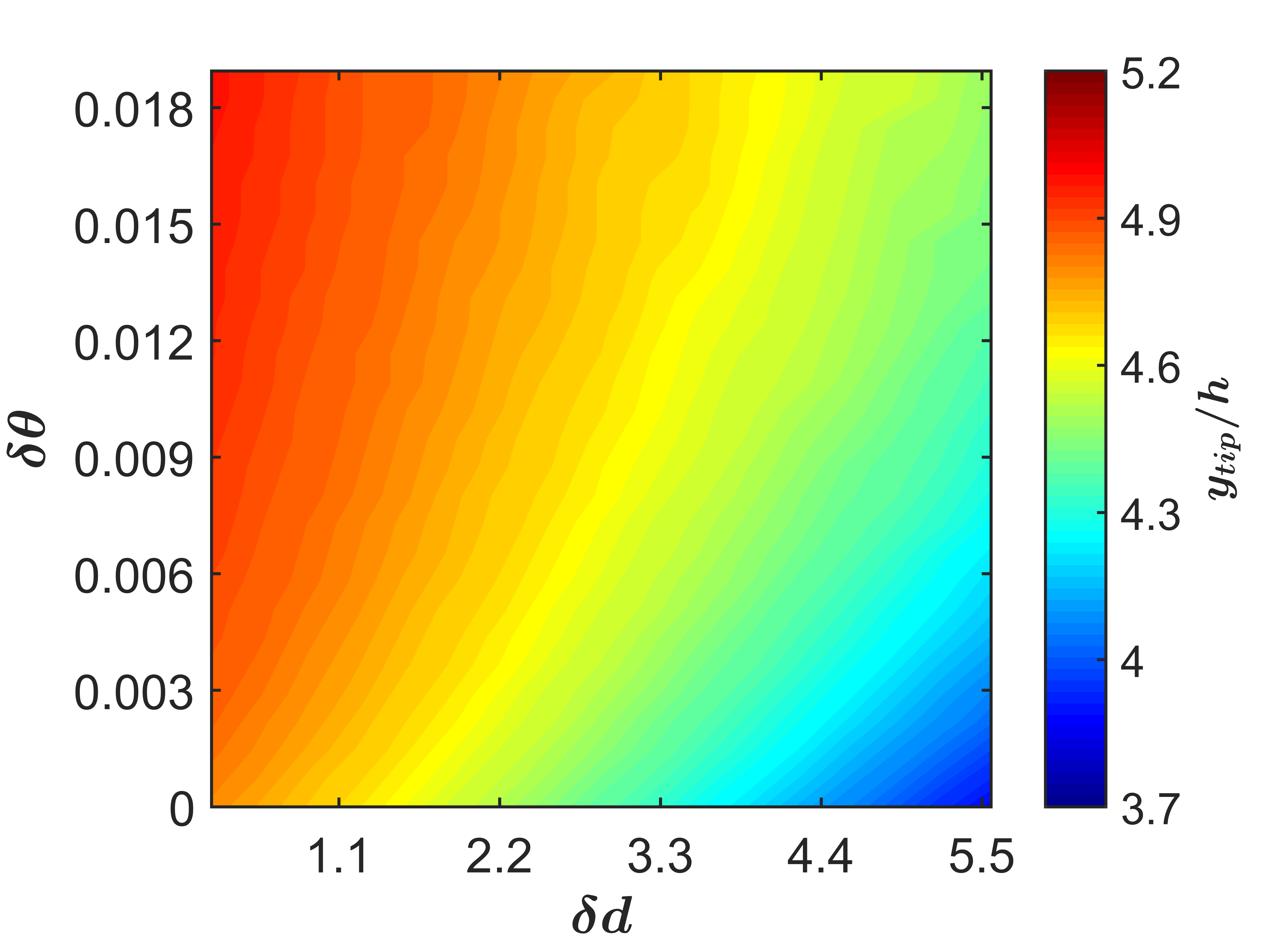}

\caption{Phase diagram of the tip deflection of FG cantilever scaly beam spanned by only positive $\delta_d$ and $\delta_\theta$ for the case of $w_0 h=1.9e-3$.}
\label{FG5} 
\end{figure}

Next, we map the landscape of stiffness over both spatial and angular orientations using a phase diagram, Fig.~\ref{FG5}. Here, we track the non-dimensional tip deflection of our FG cantilever scaly beam for positive gradients. Note that negative gradients $\delta_d < 0$ is not considered here because they do not contribute to increasing deflection with regard to uniformly scaled beam. The phase plot shows that increasing spatial gradient leads to lower deflection for any given angular distribution whereas the opposite is true for angular gradient. Thus, these two variations work in opposite directions. This feature can be used to tailor the elasticity of a cantilever beam relatively easily.

In conclusion, we explore geometrical tailorability of elasticity brought about by controlling the distribution and orientation of scales on a slender substrate.  We indicated the similarities and contrast between these structures and their conventional uniform counterparts investigated in the recent past~\cite{Ghosh1,Frank1,Hessein1}. Particularly, we showed that for the case of linear and positive gradation of scales, a lower compliance is achieved as compared to equidistance arrangement of scales. Additionally, angular and spatial functional gradation share some generality and a combination of both variations in the structure results in opposite behavior. Therefore, this study outlines the significance of surface based bioinspired strategies for making tailorable cantilever beams useful for a number of applications.


\bibliography{myreferences}
 \end{document}